# ASE-26: a curriculum for agentic software engineering as a discipline

**Mikael Gorsky** ORCID: 0009-0003-2518-9209 Holon Institute of Technology

## Abstract

The work of a professional software engineer has begun to consist, increasingly, of directing agents rather than writing code, and the empirical evidence for the shift is now several years deep. Anthropic's Economic Index puts automation at 79 per cent of Claude Code interactions [2]; Handa and colleagues at Anthropic find AI exposure for Computer Programmer tasks at approximately 75 per cent of the role's distinct activities [3]; Brynjolfsson and colleagues at Stanford's Digital Economy Lab report a 13 per cent relative decline in employment for workers aged 22 to 25 in occupations most exposed to AI [4]. The shift is also unfinished, and the academic literature on agentic software engineering converges on the finding that the missing capability is not better models but structured practitioner discipline. This paper presents ASE-26, a comprehensive undergraduate curriculum for agentic software engineering as a discipline, deposited as a citable reference on Zenodo under CC BY-ND 4.0 [12]. The paper sets out the discipline framing the curriculum rests on, the conceptual contributions it makes (most importantly, the evolutionary spiral as the operational form of the co-evolution of intent and build), the twenty-one-module structure that organises the discipline for teaching, the pedagogical commitments that follow from grading work co-produced with an agent, what graduates leave with, and how the discipline as taught is designed to outlast the specific capabilities of today's models. The position the paper takes is that the practitioner skills the industry currently lacks are precisely the skills the discipline names, and that structured undergraduate curricula in agentic software engineering are the principal mechanism by which the gap closes.

---

## 1. The work has changed; the curriculum has not

Open the chat window of any modern coding agent, whether Claude Code, Codex or Cursor, and watch a working developer ask for a feature: they write a paragraph, the agent reads it, asks two clarifying questions and then produces six hundred lines of code, a handful of tests, a commit message and a draft pull request, after which the developer reads what came back, finds a problem in the third test, asks for a fix, accepts the revised version, and commits. Eleven minutes, where three years ago the same task would have taken two hours.

This is the work now, increasingly. Anthropic's Economic Index, drawing on five hundred thousand coding-related interactions, classifies 79 per cent of Claude Code activity as automation rather than augmentation, with the human framing tasks at the start and judging output at the end while the agent does the typing in between [2]. Handa and colleagues at Anthropic, mapping Claude usage to the O*NET task taxonomy, find that AI exposure for Computer Programmers covers approximately 75 per cent of the role's distinct activities [3]. Meske and colleagues at Ruhr-Universität Bochum and Chalmers University of Technology describe what is happening at the level of the developer's experience: intent now reaches the machine through natural-language prompts and probabilistic inference, rather than through deterministic syntax [1], and the mediation layer between human intent and machine execution has been rewritten in consequence.

The change has been rapid enough to leave visible labour-market evidence. Brynjolfsson and colleagues at Stanford's Digital Economy Lab report a 13 per cent relative decline in employment for workers aged 22 to 25 in occupations most exposed to AI, even as employment for older workers in those same occupations holds steady [4]. Massenkoff and McCrory at Anthropic, working in the same vein, find a 14 per cent slowdown in hiring into entry-level positions in exposed occupations [5]. The on-ramps are narrowing before the floor falls, and a new graduate of an undergraduate programme in computer science is now entering a job market that, in three short years, has stopped asking for the work she has been trained to do.

The shift is also unfinished, and visibly so. Recent academic work on agentic software engineering converges on a single point: the technical capability of current agents has outrun the practitioner methodology required to use them well. Hassan and colleagues at Queen's University, Huawei Canada, Concordia University and the Nara Institute of Science and Technology open their SASE (Structured Agentic Software Engineering) paper with the observation that "the SE field is facing one of the most significant shifts in its history, which exposes a fundamental tension between the velocity of automation and the required rigor to build trustworthy software" [7], and they name the gap "speed vs. trust" and quantify it: a large percentage of agent-generated pull requests, on inspection, fail to meet the bar of being truly merge-ready, even as agents now produce them at industrial volumes. Otoum and Elkhalili, in their 2026 systematic literature review for *IEEE Access*, report four leading barriers to industrial adoption: trust and reliability (cited in 78 per cent of deployment-focused studies), integration with legacy systems (64 per cent), the absence of standardised evaluation methods (59 per cent) and skills gaps in development teams (52 per cent) [6]. Dong and colleagues at Google, working from ninety-one enterprise interviews, find that the missing layer is human-centred collaboration patterns rather than improved tools [9], and Wang and colleagues, in their 2025 survey of AI agentic programming, end on the same diagnosis: the technical capability is present, but the structured practitioner methodology is absent [10].

Standing in front of this shift is a generation of computer science curricula written for the work as it was, in which programming, software architecture, debugging, version control, testing and review are still taught largely as activities the human performs by hand, and in which AI tools (when they enter the syllabus at all) typically arrive as productivity aids layered on top of an unchanged practice rather than as the operational centre of a transformed discipline. The convergence of the literature suggests, in the simplest terms, that the absence of structured undergraduate curricula in agentic software engineering is the principal obstacle between today's graduating engineers and the level of practice the work now requires of them.

This paper presents ASE-26, a comprehensive curriculum for agentic software engineering as a discipline, designed for engineering students who already have foundational knowledge of computer science and some experience with software engineering; the full curriculum is deposited on Zenodo under CC BY-ND 4.0 [12]. The remaining sections set out the discipline framing the curriculum rests on, the conceptual contributions it makes, the structure that organises those contributions for teaching, the pedagogical commitments that follow from grading work co-produced with an agent, what graduates leave with, and how the discipline as taught is designed to outlast the specific capabilities of today's models. The position the paper takes is that the practitioner skills the industry currently lacks are precisely the skills the discipline names, and that structured undergraduate curricula are the principal mechanism by which the gap closes.

---

## 2. Agentic software engineering as a discipline

The opening move of ASE-26 is to insist that agentic software engineering is a discipline, and the word matters: a discipline has principles, named artefacts, recognised failure modes and standards a practitioner can be held to, and it carries from one product to the next because what it teaches is the structure of a relationship rather than the operation of any particular instrument. A casual prompter and a trained agentic software engineer can sit at the same screen, use the same model and point at the same product yet produce different things, and the difference between them is what this curriculum calls the discipline.

Hassan and colleagues at Queen's University and Huawei Canada, writing from inside the field's research community, frame this transition as the move from SE 2.0 (AI-augmented development) to SE 3.0 (agentic software engineering), and they argue that the SE 3.0 era "demands a radical reimagining of the foundational pillars of SE" [7]. Their SASE vision splits the work across two purpose-built workbenches — the Agent Command Environment (ACE), where the human orchestrates and reviews, and the Agent Execution Environment (AEE), where agents act, with structured artefacts such as the Merge-Readiness Pack (MRP) crossing between them. ASE-26 sits inside this conceptual territory, and teaches the practitioner to operate on the ACE side of the SASE split, while developing enough understanding of the AEE side to direct and verify what comes out of it.

Hoda and colleagues at Monash University push the framing further in their ICSE 2026 contribution, arguing for an agentic software engineering "beyond code" that articulates the values and vocabulary the discipline requires once it ceases to be merely a productivity story [8]. ASE-26 takes this seriously, since the curriculum's glossary is an explicit act of vocabulary construction, and the assessment design grades values (auditability, verification before trust, gatekeeper responsibility) alongside technical artefacts.

The division of labour at the centre of the discipline is straightforward to state and difficult to maintain in practice: humans frame, specify and judge, and agents execute, which means that the human's job is to equip the agent and to verify what it produces. Equipping means the right problem, the right specification, the right project context, the right tools and permissions, the right limits, and verifying means checking output against a defined standard before accepting it; each of these activities, taken seriously, is a non-trivial engineering skill, and each is what the curriculum teaches.

The division collapses in predictable ways: the human treats execution as their own responsibility, hand-edits the agent's output line by line and loses the audit trail in the process; the human delegates judgment along with execution and accepts whatever the agent produced; the human delegates framing too, lets the agent decide what the problem actually is, and ends with confident output for the wrong question. Each of these collapses is a recognisable failure mode with a recognisable kind of project debt left behind, and the discipline names them so that practitioners can recognise them, in their own work and in others'.

Four structural risks shadow the practitioner even when the division of labour holds. Skills no longer practised atrophy, so that the developer who has not written a hash table by hand in a year may lose the ability to evaluate the one the agent wrote; code produced rapidly through dialog can become opaque to the very person who directed its creation; accountability becomes diffuse when human developer, AI agent, model provider and deploying organisation each contributed to an outcome and no single one is cleanly responsible; and models reproduce the assumptions of their training corpora, including assumptions about who their users are and what design choices are defaults. None of these risks disappear with experience, but all of them can be managed by a practitioner trained to expect them.

The discipline framing answers a question the literature has been circling for several years: what is the practitioner methodology that goes alongside the technical capability? Hassan and his co-authors provide one answer in the SASE workbenches and their structured artefacts [7], Otoum and Elkhalili's review documents the absence of standardised practice [6], and Wang and colleagues catalogue the technical landscape and call for methodology [10]. ASE-26 supplies the methodology in the form a discipline takes when it is teachable: an organised body of principles, artefacts and skills, with a curriculum and an assessment design that grade students against it.

---

## 3. The conceptual model

Three commitments anchor the discipline as ASE-26 teaches it: auditability as the defining principle, the seven-part anatomy of an agentic workflow as the unit of work, and the empiricist position on intent that Section 4 will develop in detail.

Auditability is the most basic of the three: a workflow that cannot be inspected, replayed or evaluated after the fact is craft, while engineering is accountable and reproducible, and the auditability principle commits the practitioner to producing, as a routine matter, a record sufficient for someone other than the original engineer to reconstruct what was decided, why, and what happened. The artefact that makes the commitment operational is the audit trail, which is a frozen record linking intent to specification, to context, to agent trajectory, to verification. Classical software engineering struggled for decades to produce audit trails as a routine matter, because engineers wrote them by hand, after the fact, and often poorly; agentic engineering, by contrast, produces them as a byproduct of normal operation, so that the agent's reasoning sits in the chat transcript, the tool calls are logged, the artefacts are committed, and the verification gates record their own results, with the consequence that the cost of building the audit trail collapses to the cost of preserving what the workflow generated anyway, and the discipline asks for that material to be preserved, named and made navigable.

The anatomy of one workflow turn has seven parts, which run as the constituents of a single act rather than as sequential phases. Intent is the human's understanding of what is wanted, including what the project is for and who needs it to exist; specification is the precise document the agent receives, sufficient for it to make autonomous decisions correctly; context is what the engineer deliberately puts in front of the agent (the project's standards, the relevant files, the lessons learned in earlier turns); plan is the sequence of actions the agent generates given the context; execution is the agent's invocation of its tools to act on the world; verification is the gate the output must pass before it enters the project; and the audit trail records the whole thing. Each part has a corresponding module in the curriculum, and the relationships between the parts are the subject of Module 18, on the evolutionary spiral.

The empiricist position is the third commitment, and the most consequential of the three. The work of software engineering has, for most of its history, split between two philosophical camps: the rationalist holds that careful thought yields a design before any code is written and that the design can be specified completely enough to be implemented faithfully, while the empiricist holds that humans are fallible, that intent cannot be fully known before the work begins, and that design must proceed through build-evaluate-revise cycles. Frederick Brooks, in his chapter on rationalism versus empiricism in *The Design of Design*, named the axis and landed empiricist, although with respect for what upfront thought can contribute. Agentic software engineering pushes the empiricist position further still, because the agent itself behaves empirically: the output of any given prompt is a sample from a probability distribution, so reasoning about what the model should have produced is insufficient, and tests reveal what it actually produced, with the consequence that the agent must be tested, observed, corrected and tested again. Whatever upfront thinking the human contributes will land in a system that itself requires empirical handling.

The position does not dismiss upfront thought: framing the problem, writing a coherent specification and curating the context are all upstream work that pays off in fewer wasted turns. What the position holds is that the upstream work cannot complete itself before the work begins, that specification and build co-evolve, and that the mechanism by which they co-evolve is the evolutionary spiral, which Module 18 of the curriculum develops in full and which Section 4 below describes briefly.

---

## 4. The evolutionary spiral

The evolutionary spiral is the conceptual contribution at the heart of the curriculum, and it is also the synthesis on which the second pre-print in this series will turn at length. Briefly stated, it is what happens when three older ideas are put together and applied to agent-mediated work: Barry Boehm's spiral model of software development (developed at TRW Defense Systems and published in *IEEE Computer* in 1988) [13], the co-evolution of problem and solution spaces described by Mary Lou Maher and her co-authors at the University of Sydney, and later by Kees Dorst at Delft University of Technology and Nigel Cross at the Open University, in the design-research literature [14], and Frederick Brooks's contracting-point refinement of iterative process, in *The Design of Design*.

Boehm's spiral, in its original form, replaces the linear waterfall with risk-driven turns of analysis, prototyping, evaluation and planning, in which each turn produces a more definite version of the system, and the shape (which opens at the start and contracts toward the end) is what gives the model its name. Dorst and Cross, working in design rather than software, observed that the problem space and the solution space do not live in sequence in actual design practice: the designer thinks they want X, attempts a solution that resembles X, looks at the result, and realises that what was actually wanted is Y, or X minus some constraint, or X reframed for a different user. The build teaches the framer what the framer wanted, and requirements emerge from attempted solutions rather than from prior reflection. Brooks's contracting-point refinement, in *The Design of Design*, holds that inside each turn of an iterative process the work runs fast and loose, while between turns the intent is pinned down at named moments, and the architectural contracting analogy is the cleanest illustration: the client writes a programme of needs rather than a specification of solutions, the architect provides services rather than a product, conceptual designs serve as prototypes, and only at the end of the design process is a fixed-price construction contract signed.

The evolutionary spiral is what one gets when these three ideas are applied to agent-mediated software work. A project runs as a sequence of turns, inside each of which the work is fast and loose (the agent produces, the human reads, the human asks for revisions, the agent produces again) while between turns the work is pinned down (the specification is updated to reflect what the build revealed, the context file absorbs what was learned, what is done is named done, and the next turn begins from a clean restore point). Across many turns the project converges on a definite version of what was actually wanted, even though no single turn could have known the final destination.

The mechanics of one turn draw on the techniques the curriculum teaches across Part 3: intent revision (Modules 6 and 10), context update (Module 11), safety positioning (Module 12), the build itself, verification (Module 13) and review (Module 16), with multi-agent decomposition (Modules 14 and 15) and security posture (Module 17) entering as additional dimensions that shape how the turn is structured when complexity warrants. Two judgment skills make or break the rhythm, and the curriculum names both: the commit point is the decision at the end of each turn about what to lock, what goes back into the specification, and what remains open, while drift detection is the skill of recognising when the spiral is failing, whether by going in circles, by expanding scope or by losing its original intent. Both skills are learned through practice rather than through instruction, and the curriculum can name them but cannot teach them in the way it teaches, say, verification gate design.

The evolutionary spiral can be contrasted with spec-driven approaches that treat intent as fixed before the build begins; these approaches inherit decades of software engineering practice and are right about a great deal, namely that a well-specified problem produces better agent output than an underspecified one, that upfront thought reduces waste, and that contracting moments matter. Where they go wrong, in agentic engineering specifically, is in treating the specification as a closed contract, because intent moves under contact with reality, and the spec-driven framing treats this movement as a defect (as scope creep to be controlled) rather than as the central mechanism by which the framer discovers what they actually wanted. The evolutionary spiral preserves what spec-driven approaches got right and adds the recognition that the project lives through multiple turns, and the second pre-print in this series develops the contrast at full length.

The position has empirical support beyond design research, and the Lufthansa Flight 2904 accident at Warsaw in 1993 is the most arresting illustration: the autobrake software followed its specification exactly, the specification was a correct description of normal landing and a wrong description of reality (an inclined landing on a hydroplaning runway, with insufficient weight on the gear and wheels spinning below the activation threshold), and people died. No proof, no specification and no formal method can validate the original objectives against reality, because only contact with reality can do that. In agentic engineering the chat transcript is the literal mechanism through which problem and solution find each other, and a specification that did not move during a project is a specification that has not been tested against reality.

---

## 5. Twenty-one modules, four parts, one closing

The curriculum organises the discipline into twenty-one modules across four parts plus a closing module, and the structure follows the order in which a practitioner builds capability rather than the order in which the literature was published.

Part 1, *The discipline*, runs across Modules 1 through 5 and establishes what the practice is and how to read its instruments. Module 1 situates agentic software engineering historically and conceptually as the most recent step in the long arc of intent mediation, from the patch cables of the ENIAC to today's natural-language directives [1]. Module 2 develops the human role with reference to which skills erode, which remain stable and which compound as models improve, drawing on Michael Polanyi's tacit-knowing tradition for the conceptual ground and on the Otoum-Elkhalili review for the empirical evidence of industrial demand [6]. Module 3 gives the practitioner mental models of how agents actually work (the context window as the agent's whole world, tools as the agent's action space, planning as next-step generation under context, and the failure-mode taxonomy that diagnoses what went wrong when the output is bad). Module 4 names the seven-part anatomy of a workflow and the auditability principle that holds it together. Module 5 introduces the ADE (Agentic Development Environment) typology and the six-pillar architecture that lets a practitioner read any ADE, current or future, as an instance of a stable framework [7].

Part 2, *The environment*, runs across Modules 6 through 9 and equips the practitioner to direct work intelligently in a real software project. Module 6 takes problem framing seriously as a discipline with its own rigour, drawing on George Polya's *How to Solve It*, on Donald Schön's *The Reflective Practitioner*, and on Brooks's chapters on user models and constraints, to develop the practitioner's capacity to ask the right question before any agent is invoked. Module 7 teaches modern web application architecture at the level needed to direct an agent intelligently. Module 8 covers interface design and documentation under the unifying concept of legibility, which is the property that makes a system inspectable by someone who did not build it, and which becomes a control mechanism rather than a quality-of-life concern when all significant code is foreign code. Module 9 develops the architecture of intelligence in software (cognified products, where language-model intelligence is part of the user-facing value) and the agent economics that shape what such products can promise.

Part 3, *The engineering*, runs across Modules 10 through 18 and is the largest part of the curriculum. Module 10 develops specifications as the deliverable of human work, against Donald Knuth's five-criteria definition of an algorithm in *The Art of Computer Programming* as the formal benchmark a natural-language briefing approximates. Module 11 covers context engineering as a discipline of its own, with empirical support from recent research on the impact of AGENTS.md files: human-written context files improve agent performance by approximately 4 per cent, while LLM-generated context files decrease performance by 3 per cent and increase token cost by over 20 per cent [15], [16]. Module 12 covers safety, control and recovery, with Git as the primary safety layer that persists across sessions and tools. Module 13 develops verification before trust (the principle that every agent output is a hypothesis until checked, with verification theatre as a named failure mode the practitioner learns to detect). Module 14 covers multi-agent decomposition and orchestration, distinguishing implicit from explicit orchestration and introducing the five workflow patterns from Anthropic's engineering writing [17]. Module 15 develops the design of multi-agent workflows in detail. Module 16 treats code review as the skill of reading and evaluating foreign code (whether agent-generated or legacy-inherited) and introduces the MRP framework from Hassan and colleagues as the structured verification standard [7]. Module 17 covers security, risk and governance, including prompt injection as the foundational security risk, blast radius as the organising concept for permission design, and the emerging discipline of AgentOps. Module 18 closes Part 3 by returning to the evolutionary spiral and assembling the techniques of Modules 11 through 17 into the rhythm that puts them together.

Part 4, *The market*, runs across Modules 19 and 20 and equips the practitioner to navigate the world the discipline opens onto. Module 19 analyses the restructuring developer role, with reference to the AI-affects-tasks-not-jobs framing supported by Eloundou and colleagues at OpenAI, by Anthropic's Economic Index team and by the broader labour-market literature [18], [2], [4], [5], and uses the human sandwich pattern from Kieran Klaassen and Dan Shipper at *Every* as the operative metaphor for the surviving work [19], [20].

Module 20 develops a rigorous analysis of how startup economics have changed through the combined effects of agentic-coding cost compression and cognification opening new product surfaces.

The closing module, Module 21, teaches the practitioner how to keep their own mental model of the discipline fresh as the underlying technology shifts, and it is what makes the curriculum future-proof: the module teaches the practitioner to distinguish durable principles (co-evolution, intent discipline, verification before trust, human accountability, the audit trail, framing as the first design decision) from contingent capabilities (specific ADE features, current cost structures, today's failure-mode taxonomy, current context window sizes, current orchestration patterns), and to update their model of the field as model releases, ADE evolution and regulatory developments arrive.

The full text of all twenty-one modules, with reference lists and supporting materials, is deposited as the canonical reference on Zenodo [12].

---

## 6. Grading the discipline, not the artefact

The assessment design follows from a single principle: in an agentic course the artefact a student delivers is co-produced with an agent, so the grading scheme grades the discipline rather than the lines of code, and this principle takes some unpacking. It does not mean the artefact is ignored; it means that the criteria against which the artefact is judged are criteria the agent cannot, in itself, satisfy (framing, specification, context engineering, verification design, code review against a structured standard, multi-agent coordination, security and governance analysis, and the practitioner's reading of what was learned during the work), and these are the responsibilities the curriculum names as well as the parts of the work the course can reliably evaluate.

Three commitments follow from the principle, and the assessment design rests on each of them in turn.

The audit trail is gradable, which means that the student's chat transcript, their commit history, their context files, their lessons-learned notes and their verification records are submitted alongside the final artefact; without these the artefact is opaque and cannot be assessed, while with them the assessor can reconstruct what the student decided, why, and how the spiral turned. Audit trails that are present, navigable and honest score full marks on the process-traceability dimension, while audit trails that are absent, opaque or evidently retrofitted score zero, regardless of the quality of the final artefact.

The spiral is gradable as a rhythm rather than only at its endpoint, because the final state of a project rarely reveals what the student did well during the work, while the mid-project commit-point decisions, the recoveries from drift and the revisions to the specification that the build forced are the points where the student's judgment shows. The course grades multiple turns explicitly: a student who took three turns to converge on a coherent solution, with a clear audit trail showing what they learned at each commit point, scores higher than a student who arrived at a similar end state in one turn but with no record of the reasoning that got them there.

Verification is gradable as a deliverable, in the sense that a student who produced thorough verification gates that caught real failure modes scores higher than a student whose work happens to be correct but whose gates would have caught nothing; the artefact "verification gate appropriate to the failure modes this project actually faces" is what the rubric scores, with the test results as supporting evidence. Hassan and colleagues' MRP framework supplies the standard for the final deliverable, with functional completeness, sound verification, SE hygiene, rationale and communication, and full auditability all required with explicit evidence [7], and an artefact that satisfies four of the five MRP criteria is not merge-ready and is treated accordingly in the grading.

The integrity standard for an ASE course differs from the standard for a programming course in one critical respect: agent use is expected, and what the student must do is declare it, so that every submitted artefact carries a brief delegation note (which parts the student wrote themselves, which parts the agent produced under direction, and what the audit trail records). The audit trail itself is the primary check, and an artefact whose audit trail does not match its declared production process is treated as a violation. The integrity violations the course recognises are three (hidden delegation, modification of the audit trail to conceal what was delegated, and submission of artefacts that cannot be reconstructed from the audit trail), and using an agent to produce the artefact is, by itself, none of these.

The weighting of graded components is laid out in detail in the full curriculum [12]: continuous coursework comprises 50 per cent of the final grade and is collected as four portfolios across the module clusters; the project, which runs through the semester and takes a small software product across at least three full turns of the evolutionary spiral, comprises 40 per cent; and the closing artefact (a one-page document on what the student expects to revise in the curriculum within eighteen months and how they will know to revise it) comprises 10 per cent, and is what carries the meta-skill from Module 21 into the assessment itself.

---

## 7. What graduates leave with

A graduate of ASE-26 is able to do twelve things the course names explicitly, and although the list lives in full in the curriculum's learning-outcomes section [12], the summary form covers the conceptual, operational and evaluative capabilities the discipline requires.

On the conceptual side, the graduate can articulate agentic software engineering as a discipline of structured, auditable human-agent workflows, and can explain how it differs from tool skill, from casual prompting and from general LLM use, and they can apply the conceptual frame the discipline rests on (the division of labour, the auditability principle, the seven-part anatomy of an agentic workflow, the co-evolution of intent and build, and the evolutionary spiral) to any agentic project they encounter or design.

The operational capabilities are more numerous and more concrete. The graduate can frame an agentic project (problem statement, stakeholder list, testable definition of done, out-of-scope list), write a specification that approximates Knuth's five criteria, engineer the context an agent receives, apply safety infrastructure that protects against catastrophic loss, design verification gates that catch real failure modes, decompose tasks for multi-agent workflows, conduct code review against the MRP standard, identify security and governance risks specific to agentic systems, and operate a project through multiple turns of the evolutionary spiral with deliberate commit points and active drift detection.

The evaluative capabilities are the ones the curriculum spends the closing module on. The graduate can read the labour market and the entrepreneurial environment with reference to AI's effect on tasks rather than jobs, can position themselves in the bifurcation between expert-frame work and pure-implementation work, and can maintain their own mental model of the discipline as the underlying technology shifts. The last of these is the meta-capability that Module 21 develops explicitly, since the practitioner who can distinguish durable principles from contingent capabilities will remain useful when the next ADE arrives, the next model release lands or the next regulatory framework changes the rules.

The graduate is, in short, equipped to operate productively in software organisations that have adopted agentic development workflows, to evaluate agentic tools and approaches critically as they continue to evolve, and to defend the engineering standards the discipline names against the speed-first pressure that comes with agent-mediated work. The work that compounds, in the terminology of Module 2, is the work the graduate is trained to do.

The labour-market evidence for what graduates with this capability are worth is now several years deep. Dan Shipper, writing in *Every* in May 2026, argues that automation creates more expert work rather than less, because cheap competence creates demand for the differentiation that only experts can supply [19], and the market is bifurcating along this line: employers and entire industries that have absorbed the techniques of agentic engineering are paying premium for the human work that frames and judges, while employers stuck in the older mode are competing for narrowing pools of pure-implementation work. Kieran Klaassen's "bread in the AI sandwich" metaphor, also published in *Every*, captures the structure of the surviving work, in which humans frame at the start, judge at the end and the agent does the typing in between [20]. The pattern is empirical as well as metaphorical, since Anthropic's data on Claude Code shows that 79 per cent of interactions are classified as automation, but that 36 per cent of those interactions involve feedback loops where the human remains continuously in the loop [2], and the job for which the graduate is being prepared is the human side of the sandwich.

---

## 8. What lasts and what doesn't

The curriculum is built to be revised, and this is, by design, the most important thing the curriculum teaches about itself. The closing module is structured around an explicit inventory of which principles taught in the course are likely durable and which are tied to current capabilities: the durable list includes co-evolution, intent discipline, verification before trust, human accountability, the audit trail and framing as the first design decision, while the contingent list includes specific ADE features, current cost structures, today's failure-mode taxonomy, current context window sizes and current orchestration patterns. The course teaches the student to read the boundary between these two lists, and to update their model of the field as model releases, ADE evolution and regulatory developments arrive.

The historical record offers concrete evidence that the durable-versus-contingent distinction is real and observable rather than a rhetorical comfort. Donald Knuth began *The Art of Computer Programming* at Stanford in 1962 and is still publishing volumes more than six decades later, across a span during which every specific programming tool, language and machine that existed at the start has been replaced, and yet the algorithmic principles in Volume 1 remain canonical reading for serious practitioners. The principles outlasted the scaffolding by more than a working lifetime, and an undergraduate curriculum in agentic software engineering can aspire to teach principles of similar shelf life, even as the specific tools the students learn on will be obsolete by the time they apply for their second job.

The position the paper has taken throughout is that the practitioner skills the industry currently lacks are the skills the discipline names, and that structured undergraduate curricula are the principal mechanism by which the gap closes. The literature converges on the same point from several angles: Hassan and his co-authors argue for purpose-built workbenches and structured artefacts [7], Hoda and colleagues argue for an articulated values-and-vocabulary discipline beyond code [8], Dong and colleagues at Google argue for human-centred collaboration patterns [9], Otoum and Elkhalili document the skills gap empirically [6], and Wang and colleagues call for methodology to match the technical capability [10]. ASE-26 supplies the methodology in the form a discipline takes when it is taught, by providing an organised curriculum that an engineering school can deliver, an assessment design that grades the right things, and a canonical reference deposited where the field can find it.

The pre-print that follows this one will take up the conceptual contribution at the centre of the curriculum (the evolutionary spiral as the operational form of the co-evolution of intent and build) and develop it at full length, in contrast with spec-driven approaches, while a peer-reviewed paper planned with Prof. Levin will extend that contrast into a position that the broader software engineering research community can engage with directly. The Zenodo deposit is the canonical text from which all three papers draw, and the curriculum is the artefact that, if adopted, will change the discipline's transmission from a craft tradition to a teachable practice.

A discipline is what survives a generation of practitioners, and ASE-26 is one attempt to give agentic software engineering that form while the work is still young enough to take it.

---

## Foundational works referenced